\documentclass[useAMS]{tMPH2e}
\usepackage{natbib}
\title{Investigation of phase separation within the generalized Lin-Taylor model 
for a binary liquid mixture of large hexagonal and small triangular particles}
\author{J. STRE\v{C}KA\thanks{$^\ast$Corresponding
author. Email: jozef.strecka@upjs.sk\vspace{10pt}
\newline\centerline{\tiny{ {\em Molecular Physics}}}
\newline\centerline{\tiny{ISSN 0026-8976 print/ ISSN
1362-3028
 online
\textcopyright 2005 Taylor \& Francis Ltd}}
\newline\centerline{\tiny{ http://www.tandf.co.uk/journals}}
\newline \centerline{\tiny{DOI:
10.1080/002689700xxxxxxxxxxxx}}}$^\ast$, L. \v{C}ANOV\'A, and M. JA\v{S}\v{C}UR  
\\\vspace{6pt} Department of Theoretical Physics and Astrophysics, Faculty of Science, \\
P. J. \v{S}af\'arik University, Park Angelinum 9, 040 01 Ko\v{s}ice, Slovak Republic}  
\received{Received 18 July 2006}

\begin{document}
\label{firstpage} \doi{10.1080/002689700xxxxxxxxxxx}
\issn{1362-3028} \issnp{0026-8976} 

\markboth{J. Stre\v{c}ka et al.}{Investigation of phase separation within the generalized Lin-Taylor model}

\maketitle

\begin{abstract}
The generalized Lin-Taylor model defined on the hexagonal lattice is used to investigate the phase separation 
in an asymmetric binary liquid mixture consisting of large A (hexagons) and small B (triangles) particles. 
By considering interaction energies between A-A and A-B pairs of particles that occupy nearest-neighbour cells of the hexagonal lattice, we have derived an exact solution for the considered model system having established a mapping correspondence with the two-dimensional Ising model on its dual triangular lattice. Altogether, six different types of coexistence curves including those with reentrant miscibility regions (i.e. closed-loop coexistence curves) were found in dependence on the relative strength between both 
coupling constants.  
\end{abstract}

\section{Introduction}
\label{intro}

Phase equilibria of binary liquid mixtures are subject of immense research interest since closed loops 
of immiscibility were experimentally observed in the nicotine-water mixture \cite{Hud04} and later on, 
in several aqueous solutions of the 1-, 2-, 3- and 4-methylpiperidine \cite{Fla08a, Fla08b, Fla09}, 
$\alpha$- and $\beta$-picoline \cite{And52,Cox52,Gar76}, 2,4- and 2,6-dimethylpyridine \cite{Fla09, Jon21}, 
the glycerol-guaiacol mixture \cite{Ewe23}, and also in many polymeric blends \cite{Nar94}. It is noteworthy that an existence of the closed-loop coexistence curves with both upper ($T_{\rm U}$) as well as lower ($T_{\rm L}$) critical solution temperatures evidently contradicts with intuitive expectations based on thermodynamical description of demixing, which predicts, on the contrary, a bell-shaped coexistence curve terminating at a single $T_{\rm U}$ instead of having an island of immiscibility. Early explanation of this remarkable and rather unexpected phenomenon has been suggested by Hirschfelder, Stevenson and Eyring \cite{Hir37} who associated an appearance of the reentrant miscibility with a presence of highly orientation-dependent forces, such as hydrogen bonding, 
which becomes rather inefficient above a certain critical temperature. In addition, if directional bonding occurs between like as well as unlike particles and the sum of interaction energies for pairs of like particles is simultaneously greater than for pairs of unlike particles, the binary mixture may even phase separate at low temperatures to yield an usual bell-shaped coexistence curve with the low-temperature $T_{\rm U}$ below an additional closed loop. Although one meets with such an intriguing situation rather rarely because
freezing transitions usually camouflage the lowest $T_{\rm U}$, the aqueous solution of 2-butanol provides 
a striking example of the binary mixture with the closed loop above the bell-shaped coexistence curve \cite{Dol08,Mor75,Sor88,Mon93}.

So far, different lattice models of liquid mixtures have been proposed and dealt with in order to 
bring an insight into a reentrant miscibility phenomenon \cite{Nar94}. Even though all lattice 
models provide somewhat oversimplified picture of the liquid mixtures, some of the lattice models 
have already proved their worth due to their capability to explain the reentrant miscibility of hydrogen-bonded liquid mixtures \cite{Whe80,Wal83, Vey90}. Most of theoretical predictions for 
a closed-loop formation were indeed based on the lattice-statistical models naturally describing 
the directional character of highly orientationally dependent forces and moreover, the lattice 
models can be rather easily treated using relatively precise Bethe-Guggenheim quasichemical 
approximation \cite{Gug52,Bar53,Bod84} or Migdal-Kadanoff approach formulated on the basis 
of renormalization group theory \cite{Wal80,Wal82,Gol83a,Gol83b,Wal87}.
Besides, there even exist few exactly solvable lattice models, such as Frenkel-Louis model \cite{Fre92} 
or Lin-Taylor model (LTM) \cite{Lin94a,Lin94b}, which give qualitatively correct results for a phase separation without being affected by any approximation. It is noteworthy, however, that more complete understanding of the reentrant miscibility has been achieved by exploring off-lattice continuum 
models, which can be treated within the perturbation theory of Wertheim \cite{Jac91,Gar98a,Gar98b} 
or the Gibbs-ensemble Monte-Carlo simulation technique \cite{Dav99,Dav00}.

In the present article, we shall treat one particular example of LTM recently 
remarkably generalized by Romero-Enrique and co-workers \cite{Rom97,Rom98a,Rom98b,Rom99,Rom03}. 
Among other matters, the generalized LTM has proved its usefulness in elucidating the reentrant 
miscibility as it has provided a plausible explanation of its microscopic origin. Even if the 
procedure worked out previously by Romero-Enrique \textit{et al}. \cite{Rom98a} is rather general 
and contains our version of the model only as a special case, the numerical results reported on 
previously were mostly restricted to LTM on a square lattice, which is a rather exceptional 
case of the self-dual lattice. In this respect, we shall therefore examine here the phase separation
in the binary mixture of large hexagonal particles and small triangular ones, which are allowed to occupy basic unit cells of the hexagonal lattice. As we shall see hereafter, this model also exhibits the same substantial diversity of phase diagrams as remarked by Romero-Enrique and co-workers \cite{Rom98a} and 
all available scenarios for coexistence curves including those with a closed loop of immiscibility will 
be indeed confirmed. 

The organization of this paper is as follows. The foundations of the generalized LTM on the hexagonal 
lattice and basic steps of the exact method are given in the section \ref{model}. Section \ref{result}
deals with the most interesting numerical results obtained for the phase diagrams and coexistence curves.
Finally, some conclusions are mentioned in the section \ref{conclusion}.

\section{Generalized Lin-Taylor model (LTM)}
\label{model}

Let us begin by recalling foundations of the generalized LTM on the hexagonal lattice. Assuming 
that the binary mixture consists of two kind of species, the large hexagonal particles A and the
small triangular ones B, LTM on the hexagonal lattice can be defined through the following rules: 
(i) each hexagonal unit cell consists of six triangular sub-units; (ii) none of A-A and A-B pairs can 
occupy the same hexagon; (iii) the small particles B can occupy the cells left empty by the large particles A; (iv) each triangular sub-unit can be occupied by at most one particle B; (v) pairwise interactions $\varepsilon_{\rm AA}$ and $\varepsilon_{\rm AB}$ exist between the nearest-neighbouring A-A and A-B pairs 
sharing a common edge (Fig. 1). Next, we shall refer to $\mu_{\rm A}$ and $\mu_{\rm B}$ as to the chemical potentials relative to the particles A and B, respectively. For further convenience, let us also define fugacities of both the particles by 
\begin{eqnarray}
z_{\rm A,B} = \exp(\beta \mu_{\rm A,B}), 
\label{fuga}
\end{eqnarray}
where $\beta = 1/ (k_{\rm B} T)$, $k_{\rm B}$ is Boltzmann's constant and $T$ absolute temperature.
The grand-canonical partition function of this binary liquid mixture can be written in the form of 
a sum taken over all occupation configurations $\{ n_i \}$ available for a given system
\begin{eqnarray}
\Xi = \sum_{\{ n_i \}} \exp \Bigl(\beta \mu_{\rm A} \sum_{i} n_i 
    - \beta \varepsilon_{\rm AA} \sum_{(i,j)} n_i n_j \Bigr) \, \Xi_{\rm B} (\{ n_i \}). 
\label{gcpf}
\end{eqnarray} 
Above, $n_i=1,0$ labels occupation number of the basic unit cell (hexagon), whereas $n_i = 1$ if $i$th 
hexagon is occupied by the large particle A, otherwise $n_i = 0$. The summation $\sum_{i}$ runs over 
all basic unit cells of the hexagonal lattice and the other one $\sum_{(i,j)}$ is performed over all 
pairs of nearest-neighbour unit cells. If the configuration of large hexagonal particles A is fixed, 
the grand-canonical partition function $\Xi_{\rm B} (\{ n_i \})$ of the small particles B can be 
expressed as follows
\begin{eqnarray}
\Xi_{\rm B} (\{ n_i \}) = (1 + z_{\rm B})^{M_{\rm F}}
(1 + z_{\rm B} {\rm e}^{-\beta \varepsilon_{\rm AB}})^{M_{\rm I}}. 
\label{gcpfb}
\end{eqnarray} 
Here, $M_{\rm F}$ ($M_{\rm I}$) denotes the total number of triangular sub-units available for occupancy 
to the small particles B, which have (not) a common edge with the large particles A. When taking into 
account the lattice geometry, both the numbers can easily be evaluated in terms of the occupation numbers 
\begin{eqnarray}
{M_{\rm F}} &=& 6 \sum_{i} n_i - 2 \sum_{(i,j)} n_i n_j, \label{numb1} \\  
{M_{\rm I}} &=& 6N - 12 \sum_{i} n_i + 2 \sum_{(i,j)} n_i n_j, 
\label{numb2}
\end{eqnarray} 
where $N$ labels the total number of the lattice cells (hexagons).
At this stage, the grand-canonical partition function of the binary mixture can be mapped onto 
an one-component lattice-gas model for the species A by substituting (\ref{gcpfb}), (\ref{numb1}) and (\ref{numb2}) into the relation (\ref{gcpf}) and thus, performing a summation over the occupation 
variables relative to the species B
\begin{eqnarray}
\Xi &=& (1 + z_{\rm B})^{6N} \sum_{\{ n_i \}} \exp \Bigl \{ [-\beta \varepsilon_{\rm AA} 
- 2 \ln(1 + z_{\rm B} {\rm e}^{-\beta \varepsilon_{\rm AB}}) + 2 \ln(1 + z_{\rm B})] \sum_{(i,j)} n_i n_j \Bigr \} \nonumber \\
&& \mbox{x} \; \exp \Bigl \{[\beta \mu_{\rm A} + 6 \ln(1 + z_{\rm B} {\rm e}^{-\beta \varepsilon_{\rm AB}}) 
- 12 \ln(1 + z_{\rm B})] \sum_{i} n_i \Bigr \}. 
\label{lg}
\end{eqnarray} 
The equality (\ref{lg}) proves that the grand-canonical partition function can be calculated from 
the equivalent lattice gas model with the renormalized pair interaction and shifted chemical potential. 
Next, the transformation between the occupation number ($n_i = 1,0$) and Ising spin variable ($s_i= \pm 1$)
\begin{eqnarray}
n_i = (1 + s_i)/2, 
\label{mt}	
\end{eqnarray}
establishes a mapping relationship between the model under investigation and the spin-$\frac12$ Ising 
model on the dual triangular lattice. As a consequence of this, the grand-canonical partition function
can be expressed through the partition function of the Ising model on the triangular lattice  
\begin{eqnarray}
\Xi = [z_{\rm A}^2 {\rm e}^{-3 \beta \varepsilon_{\rm AA}} (1 + z_{\rm B})^{6}
        (1 + z_{\rm B} {\rm e}^{-\beta \varepsilon_{\rm AB}})^{6}]^{\frac{N}{4}} Z_{\rm triang} (R, H). 
\label{map}
\end{eqnarray} 
Note that the mapping parameters $R$ and $H$ stand for the effective coupling constant and effective magnetic 
field in the associated spin-$\frac12$ Ising model on the triangular lattice
\begin{eqnarray}
R &=& \frac{1}{4} \ln \Bigl( {\rm e}^{-\beta \varepsilon_{\rm AA}} 
\frac{(1 + z_{\rm B})^2}{(1 + z_{\rm B}{\rm e}^{-\beta \varepsilon_{\rm AB}})^2} \Bigr) 
\label{par1} \\
H &=& \frac{1}{2} \ln \Bigl( {\rm e}^{- 3 \beta \varepsilon_{\rm AA}} 
\frac{z_{\rm A}}{(1 + z_{\rm B})^6} \Bigr).
\label{par2}	
\end{eqnarray}
It is worthwhile to remark that only the requirement of zero effective field ($H=0$) in the 
equivalent Ising model ensures a possible existence of the phase separation in the considered LTM.
As a result, the necessary (but not sufficient) condition allocating the coexistence 
region between two phases of different composition relates two fugacities of the large and small 
particles through $z_{\rm A} = {\rm e}^{3 \beta \varepsilon_{\rm AA}} (1 + z_{\rm B})^6$. 

Notice that the mapping relation (\ref{map}) represents a central result of our calculation 
as it formally completes an exact solution for the grand-canonical partition function $\Xi$ 
with regard to the known exact result of the partition function of the triangular Ising model \cite{Hou50,Tem50}. Besides, this relationship can readily be utilized for calculating the number 
density of the large ($n_{\rm A}$) and small ($n_{\rm B}$) particles by relating them to the 
magnetization per site ($m_{\rm triang}$) and the internal energy per site ($u_{\rm triang}$) 
of the corresponding Ising model on the triangular lattice \cite{Hou50,Tem50,Pot52}. 
After straightforward calculation one actually finds
\begin{eqnarray}
n_{\rm A}	= \frac{1 + m_{\rm triang}}{2}, \quad
n_{\rm B}	= \frac{3}{2}(C_1 + C_2) - 3 C_1 m_{\rm triang} - \frac{1}{2} (C_1 - C_2) u_{\rm triang}, 
\label{dens}
\end{eqnarray}
where the coefficients $C_1$ and $C_2$ are defined as follows
\begin{eqnarray}
C_1 = \frac{z_{\rm B}}{1 + z_{\rm B}}, \quad 
C_2 = \frac{z_{\rm B}}{{\rm e}^{\beta \varepsilon_{\rm AB}} + z_{\rm B}}.
\label{koef}
\end{eqnarray}
The composition of the binary mixture is subsequently unambiguously determined 
by the molar fraction of the large particles A defined by means of
\begin{eqnarray}
X_{\rm A} = \frac{n_{\rm A}}{n_{\rm A} + n_{\rm B}}.
\label{mol}	
\end{eqnarray}
Finally, the pressure of the binary mixture can be calculated 
from the grand potential $\Omega$ by the use of
\begin{eqnarray}
p^{*} = - \frac{\Omega}{N \sigma} = \frac{k_{\rm B} T \ln \Xi}{N \sigma},
\label{press}	
\end{eqnarray}
where $\sigma$ labels the area of basic unit cell (hexagon). For simplicity, 
we shall use throughout the rest of this paper the renormalized quantity 
$p = p^{*} N \sigma$ as a measure of the pressure. 

\subsection{Coexistence region} 

In order to describe the phase separation into two phases of different composition, 
let us firstly express the fugacity and pressure as a function of temperature, 
the interaction parameters $\varepsilon_{\rm AA}$, $\varepsilon_{\rm AB}$ and the effective coupling $R$.
According to the equation (\ref{par1}), the fugacity of small particles $B$ can be calculated from
\begin{eqnarray}
z_{\rm B} = \frac{{\rm e}^{2R} - {\rm e}^{-\beta \varepsilon_{\rm AA}/2}}
                 {{\rm e}^{-\beta \varepsilon_{\rm AA}/2} - {\rm e}^{2R - \beta \varepsilon_{\rm AB}}}.
\label{fugab}
\end{eqnarray}
Furthermore, substituting (\ref{fugab}) and (\ref{map}) into the equation (\ref{press}) 
gives the following expression for pressure 
\begin{eqnarray}
p = k_{\rm B} T \Bigl \{ -3R  + 6 \ln \Bigl[ \frac{1 - {\rm e}^{-\beta \varepsilon_{\rm AB}}}
{{\rm e}^{-2R - \beta \varepsilon_{\rm AA}/2} - {\rm e}^{- \beta \varepsilon_{\rm AB}}} \Bigr]
+ \frac{1}{N} \ln Z_{\rm triang} \Bigr \}.
\label{pressure}	
\end{eqnarray}
Experimental conditions under which coexistence curves are usually observed meet the requirement of constant pressure. In this respect, the equation (\ref{pressure}) can serve for determining the effective coupling parameter $R$ by selecting $T$, $\varepsilon_{\rm AA}$ and $\varepsilon_{\rm AB}$ at fixed value 
of pressure. A subsequent substitution of the effective coupling $R$ into the equation (\ref{fugab}) then
enables a simple calculation of the fugacities (the fugacity of large particles is connected to the one 
of small particles due to an unconditional validity of the zero-field condition). With all this in mind, 
the composition of binary liquid mixture can be consequently calculated using the set of equations (\ref{dens})-(\ref{mol}).

\subsection{Critical behaviour}

According to the mapping relation (\ref{map}), the binary mixture becomes critical if,
and only if, the effective coupling parameter of the associated Ising model approaches its 
critical value. Owing to this fact, the critical points can be found from this simple equation
\begin{eqnarray}
p_{\rm c} = k_{\rm B} T_{\rm c} \Bigl \{ 0.055627 + 
6 \ln \Bigl[ \frac{\sqrt{3} (1 - {\rm e}^{-\beta_{\rm c} \varepsilon_{\rm AB}})}{{\rm e}^{- \beta_{\rm c} 
\varepsilon_{\rm AA}/2} - \sqrt{3} {\rm e}^{- \beta_{\rm c} \varepsilon_{\rm AB}}} \Bigr] \Bigr \}, 
\label{crit}
\end{eqnarray}
where $\beta_{\rm c} = 1/(k_{\rm B} T_{\rm c})$, $p_{\rm c}$ and $T_{\rm c}$ label the critical pressure and critical temperature. The molar fraction of large particles at a critical point can be consecutively simplified to
\begin{eqnarray}
X_{\rm A}^{\rm c} = \Bigl \{ 1 + 5 \frac{\sqrt{3} - {\rm e}^{-\beta_{\rm c} \varepsilon_{\rm AA}/2}}
{\sqrt{3}(1 - {\rm e}^{-\beta_{\rm c} \varepsilon_{\rm AB}})} + \frac{\sqrt{3} {\rm e}^{\beta_{\rm c} 
\varepsilon_{\rm AA}/2} - 1}{{\rm e}^{\beta_{\rm c} \varepsilon_{\rm AB}} - 1} \Bigr \}^{-1}.
\label{molc}	
\end{eqnarray}
 
\section{Results and Discussion}
\label{result}

In this part, we shall briefly discuss the most interesting results obtained for the phase diagrams 
and coexistence curves of the generalized LTM on the hexagonal lattice. It is worthy to mention 
that the displayed coexistence curves are obtained by cutting concentration-temperature slices at 
fixed pressure, i.e. a situation which fully corresponds experimental conditions by performing studies 
of the phase separation. It directly follows from the equation (\ref{fugab}) that the phase separation 
into A-rich and B-rich phases might in principle appear either if 
$\varepsilon_{\rm AA}>0$, $\varepsilon_{\rm AB}/\varepsilon_{\rm AA}>0.5$, 
or $\varepsilon_{\rm AA}<0$ (arbitrary $\varepsilon_{\rm AB}$). 
In what follows, we shall treat those special cases in several sub-sections.  

\subsection{Repulsive interaction $\varepsilon_{\rm AA}>0$} 

If the pair interaction between the particles A is repulsive, the coexistence then possibly occurs just 
as the interaction between the A-B pairs is likewise repulsive and simultaneously, the repulsion energy $\varepsilon_{\rm AB}$ is stronger than a half of the repulsion energy $\varepsilon_{\rm AA}$, i.e. $\varepsilon_{\rm AB}/\varepsilon_{\rm AA} > 0.5$. The phase diagram for this particular case is depicted in 
Fig. 2a in the form of $T_{\rm c}-p_{\rm c}$ dependence. The region inherent to the coexistence of the A-rich and B-rich phases can be located above the displayed curves, while below them both the components become perfectly miscible. It is quite obvious from Fig. 2a that the phase separation occurs only above a certain boundary pressure $p_{\rm b}/\varepsilon_{\rm AA}=3$, which is needed for overcoming the repulsive force between the pairs of particles A. On the other hand, the mixture becomes perfectly miscible independently 
of the ratio $\varepsilon_{\rm AB}/\varepsilon_{\rm AA}$ below this pressure value. It can be also readily understood from Fig. 2a that the stronger the repulsion energy between the A-B pairs, which means, the stronger the ratio 
$\varepsilon_{\rm AB}/\varepsilon_{\rm AA}$, the higher the critical temperature at which the phase coexistence disappears at a given pressure. Next, the lines depicted in Fig. 3a illustrate changes of the critical concentration $X_{\rm A}^{\rm c}$ along the critical lines from Fig. 2a. As one can see, the critical concentration achieves $X_{\rm A}^{\rm c} = \frac16$ regardless of $\varepsilon_{\rm AB}/\varepsilon_{\rm AA}$ 
when approaching the lowest pressure $p_{\rm b}$ at which the phase separation appears. 
For completeness, three typical coexistence curves are shown in Fig. 4a for $\varepsilon_{\rm AB}/\varepsilon_{\rm AA} = 1$ fixed and several values of pressure. As it can be clearly seen, 
one finds the typical bell-shaped coexistence curves irrespective of the pressure strength.   
   
\subsection{Attractive interaction $\varepsilon_{\rm AA}<0$} 

Now, let us examine the case when the pairwise interaction between the particles A is attractive. 
In such a case one encounters much more complex phase diagrams and also much richer critical behaviour. 
For better orientation, we have divided our discussion into several parts dealing with some typical cases. 

\subsubsection{$\varepsilon_{\rm AB}>0$} 
In this case, the repulsive interaction between the A-B pairs favours the phase separation into A-rich 
and B-rich phases and as a consequence of this, one finds a quite similar phase diagram and coexistence 
curves as discussed in above (compare Fig. 2b with 2a). The most significant difference consists in an appearance of a special critical end point with coordinates 
$[k_{\rm B} T_{\rm c}^{*}/|\varepsilon_{\rm AA}|, p_{\rm c}^{*}/|\varepsilon_{\rm AA}|] = [0.910,0.051]$,
which is labelled by a star symbol in Figs. 2b)-d). While for pressures greater than $p_{\rm c}^{*}$ 
the critical temperature monotonically increases as pressure rises, below the pressure $p_{\rm c}^{*}$ 
there is a coexistence but no criticality. This observation would suggest that the critical end point terminates at a liquid-vapour coexistence line of the pure A component. Actually, the mixture becomes pure A 
before it turns critical for any pressure lower than $p_{\rm c}^{*}$. This fact can be clearly seen also 
from Fig. 3b, where the critical concentration is plotted against the critical temperature and 
all lines start from the same critical temperature of the pure A component ($X_{\rm A}^{\rm c} = 1$). 
For illustration, we depict in Fig. 4b three typical coexistence curves that obviously exhibit 
much more pronounced asymmetry than those discussed formerly for the case with the repulsive interaction 
$\varepsilon_{\rm AA}$. Apparently, this asymmetry is the more evident, the lower and the closer is 
pressure selected to its critical value $p_{\rm c}^{*}$, since the initially B-rich phase looses at low pressures much more rapidly its B component than the A-rich phase is enriched by the particles B. 
 
\subsubsection{$-0.5 < \varepsilon_{\rm AB}/|\varepsilon_{\rm AA}| < 0$} 
Contrary to the aforedescribed behaviour, it is easy to understand from Fig. 2c that a weak attractive 
force between the A-B pairs leads to a monotonous decrease of the critical temperature when increasing 
the pressure strength. The standard bell-shaped coexistence curves, which are plotted in Fig. 4c for 
one particular ratio $\varepsilon_{\rm AB}/|\varepsilon_{\rm AA}| = -0.3$ and several values of pressure,
provide a strong support for this statement. According to these plots and also dependences drawn in Fig. 3c, 
a suppression of the critical concentration $X_{\rm A}^{\rm c}$ in response to a pressure strengthening  
is observed due to a change of the character of the interaction energy $\varepsilon_{\rm AB}$. This rather easily understandable behaviour survives unless the ratio between both coupling constants does not reach 
the value $\varepsilon_{\rm AB}/|\varepsilon_{\rm AA}| = -0.431$. Within the interval 
$-0.5 < \varepsilon_{\rm AB}/|\varepsilon_{\rm AA}| < -0.431$, however, an outstanding non-monotonous dependence of the critical temperature on pressure can be detected. Assuming for instance that 
$\varepsilon_{\rm AB}/|\varepsilon_{\rm AA}| = -0.45$ is kept constant, the $T_{\rm c}-p_{\rm c}$ dependence 
can be characterized by one local minimum 
$[k_{\rm B} T_{\rm c}^{\rm min}/|\varepsilon_{\rm AA}|, p^{\rm min}_{\rm c}/|\varepsilon_{\rm AA}|] 
= [0.146,1.436]$ 
and one local maximum 
$[k_{\rm B} T_{\rm c}^{\rm max}/|\varepsilon_{\rm AA}|, p^{\rm max}_{\rm c}/|\varepsilon_{\rm AA}|] = [0.353,1.643]$. 
While below $p^{\rm min}_{\rm c}$ or above $p^{\rm max}_{\rm c}$ the usual bell-shaped coexistence curves 
with single $T_{\rm U}$ should be expected, the coexistence curves with three consecutive critical points ($T_{\rm U}$, $T_{\rm L}$, and $T_{\rm U}$) and reentrant miscibility should emerge for any pressure from inside the interval bounded by $p^{\rm min}_{\rm c}$ and $p^{\rm max}_{\rm c}$. Fig. 4d illustrates such example of a closed loop above a bell-shaped coexistence curve with in total three critical points obtained at pressure
$p/|\varepsilon_{\rm AA}|=1.5$, 
as well as, the usual bell-shaped coexistence curve to emerge when $p/|\varepsilon_{\rm AA}|=0.5$. 
If pressure is selected sufficiently close but still below $p^{\rm min}_{\rm c}$, the 'hour-glass' coexistence curve can be even detected; temperature induces after initial increase of mutual solubility its decrement until a repeated increase of solubility near $T_{\rm U}$ re-appears, as it is clearly depicted for the particular case $p/|\varepsilon_{\rm AA}| = 1.0$. It is worthwhile to remark that the aforementioned 
behaviour is completely consistent with experimental observations of the pressure effect on the 
coexistence curves of the aqueous solution of 2-butanol \cite{Mor75,Sor88,Mon93}. 

Even more involved situation emerges if pressure is selected directly equal to $p^{\rm min}_{\rm c}$ 
or $p^{\rm max}_{\rm c}$. When the upper bound $p^{\rm max}_{\rm c}$ is selected, then $T_{\rm U}$ and 
$T_{\rm L}$ incident to an island of immiscibility coalesce at so-called double critical points which are shown as circles. On the other hand, the low-temperature $T_{\rm U}$ of bell-shaped curve merges together 
with $T_{\rm L}$ of the closed loop to yield a critical double point (which is marked as diamond) by 
selecting the lower bound $p^{\rm min}_{\rm c}$ for pressure. Notice that the double critical points as well as the critical double point can be characterized by a doubling of the critical exponents as it has been proved previously \cite{Rom98b}. However, the most peculiar critical point 
$[k_{\rm B} T_{\rm c}^{\rm ip}/|\varepsilon_{\rm AA}|, p^{\rm ip}_{\rm c}/|\varepsilon_{\rm AA}|] 
= [0.265,1.862]$ 
appears by considering following value for the ratio between both coupling constants
$\varepsilon_{\rm AB}/|\varepsilon_{\rm AA}| = -0.431$. 
In this case all three critical temperatures coalesce simultaneously at so-called critical 
inflection point, which can be characterized by a tripling of the critical exponents \cite{Rom98b}. 

\subsubsection{$\varepsilon_{\rm AA}<0$, $\varepsilon_{\rm AB}/|\varepsilon_{\rm AA}|<-0.5$}   

Finally, we shall discuss the coexistence phenomenon for the case when the attractive force 
between the A-B pairs is stronger than a half of the attractive force between the A-A pairs, which means, 
$\varepsilon_{\rm AB}/|\varepsilon_{\rm AA}|<-0.5$. Under these circumstances, the coexistence 
region can be located below the curves displayed in Fig. 2d. It is quite obvious from this figure 
that there appears a closed loop of immiscibility whenever pressure is selected above $p_{\rm c}^{*}$ 
but below the value $p^{\rm max}_{\rm c}$ corresponding to the double critical point. Namely, 
the miscibility gap $\Delta = T_{\rm U} - T_{\rm L}$ gradually decreases upon pressure strengthening 
until it vanishes at the double critical points where $T_{\rm U}$ and $T_{\rm L}$ merge together. 
This scenario can be clearly seen from Fig. 4e, where the closed-loop coexistence curves are plotted 
for three different pressures and the ratio $\varepsilon_{\rm AB}/|\varepsilon_{\rm AA}|=-1$. 
It is quite apparent from this figure that the stronger the pressure, the smaller the miscibility gap (coexistence region). Note furthermore that there even exists another possible scenario to have 
a coexistence without criticality at $ T_{\rm U}$ when selecting pressure below 
$p_{\rm c}^{*}/|\varepsilon_{\rm AA}|=0.051$. In such a case, the temperature induces a phase separation 
at $T_{\rm L}$, nevertheless, both separated phases become pure A before the mixture turns 
critical and hence, one observes two pure phases of particles A in coexistence that merely differ 
in their densities. This observation can be interpreted as a liquid-vapour phase separation of 
the pure A component, which takes place because the component B vapourises prior to achieving 
the critical temperature when pressure  is selected below its critical value $p_{\rm c}^{*}$.

\section{Conclusion}
\label{conclusion}

The present article deals with the investigation of phase separation in an asymmetric binary 
liquid mixture of large hexagonal and small triangular particles described by means of the
generalized LTM on the hexagonal lattice. Despite its simplicity and a certain oversimplification, 
this model is relevant as it provides deeper understanding of the phase separation phenomenon without 
making any further approximation to the results obtained. In addition, this model proves an existence 
of the closed loops of immiscibility occurring under certain conditions either separately, 
or above the standard bell-shaped coexistence curves. The closed-loop coexistence curve bounded 
by two critical points $T_{\rm U}$ and $T_{\rm L}$ indicates such a kind of the reentrant miscibility, 
where $T_{\rm L}$ determines an upper critical temperature under which two components become repeatedly perfectly miscible. On the other hand, the reentrant miscibility can also be found in the closed loop plus bell-shaped coexistence curve with in total three critical points; the region of reentrant miscibility 
then occurs in between  $T_{\rm L}$ of the closed loop and $T_{\rm U}$ of the bell-shaped coexistence curve. Altogether, six possible scenarios for the phase separation were illustrated with the help of exact results for the phase diagrams and coexistence curves: the standard bell-shaped curve with $T_{\rm U}$, 
the bell-shaped curve without $T_{\rm U}$, the bell-shaped curve plus closed loop with two $T_{\rm U}$ 
and one $T_{\rm L}$, the closed-loop curve with $T_{\rm U}$ and $T_{\rm L}$, the 'hour-glass' 
curve with $T_{\rm U}$, and the closed loop with $T_{\rm L}$ but without $T_{\rm U}$. 

The main objective of the present work was to provide an eventual confirmation of the aforementioned coexistence scenarios, which were originally envisaged by Romero-Enrique and co-workers after introducing and exploring the generalized LTM \cite{Rom98a}. It should be pointed out, however, that the majority of numerical results reported on hitherto were mostly restricted to the particular case of the generalized LTM on the square lattice. From this point of view, the investigation of LTM on the hexagonal lattice is of a particular importance, because this model even has more obvious relevance to the phase separation of real binary liquid mixtures which might consist of molecules with a hexagonal symmetry (like benzene, cyclohexane and a large number of their structural and heterocyclic derivatives) and smaller non-linear molecules of the solvent 
(like acetone, isopropanol, ethyleneoxide, ethers, dimethylsulfoxide, etc.). Our next effort is to provide 
a further extension to the model under consideration to account also for multiparticle interactions and 
to elucidate a role of the size of solvent on the mutual solubility.

\vspace{2cm}

\textbf{Figure Captions:}
\begin{itemize}
\item [Fig. 1]
One among possible configurations of particles within LTM on the hexagonal lattice. 
Large grey hexagons denote lattice positions of the particles A, small black 
triangles positions of the particles B. The interactions are considered only 
between A-A and A-B pairs of nearest neighbours, which share a common edge (point contacts are neglected). 

\item [Fig. 2]
Phase diagrams in the form of $T_{\rm c} - p_{\rm c}$ dependence for four different cases considered:
a) $\varepsilon_{\rm AA}>0$, $\varepsilon_{\rm AB}/\varepsilon_{\rm AA}>0.5$; 
b) $\varepsilon_{\rm AA}<0$, $\varepsilon_{\rm AB}/|\varepsilon_{\rm AA}|>0$;
c) $\varepsilon_{\rm AA}<0$, $-0.5 < \varepsilon_{\rm AB}/|\varepsilon_{\rm AA}|<0$;
d) $\varepsilon_{\rm AA}<0$, $\varepsilon_{\rm AB}/|\varepsilon_{\rm AA}|<-0.5$.
The star symbol denotes a special critical end point terminating on a liquid-vapour coexistence line 
of the pure A component, the circles and diamond label the double critical points and critical double 
point, respectively. For clarity, the points at which coexistence of two different pure A phases disappears prior to achieving the critical point is shown as a broken line only in Fig. 2d) for the particular ratio 
$\varepsilon_{\rm AB}/|\varepsilon_{\rm AA}|=-1$. 

\item [Fig. 3]
The changes of critical concentration $X_{\rm A}^{\rm c}$ along the critical lines displayed in Fig. 1.
The star symbol denotes the critical end point terminating on a liquid-vapour coexistence line 
of the pure A component. 

\item [Fig. 4]
Some typical examples of the coexistence curves displayed in the form of composition vs. temperature  dependence for: a) $\varepsilon_{\rm AA}>0$, $\varepsilon_{\rm AA}/\varepsilon_{\rm AB} = 1$, 
b) $\varepsilon_{\rm AA}<0$, $\varepsilon_{\rm AA}/|\varepsilon_{\rm AB}| = 1$, c) $\varepsilon_{\rm AA}<0$, $\varepsilon_{\rm AA}/|\varepsilon_{\rm AB}| = -0.3$, d) $\varepsilon_{\rm AA}<0$, 
$\varepsilon_{\rm AA}/|\varepsilon_{\rm AB}| = -0.45$, e)-f) $\varepsilon_{\rm AA}<0$, 
$\varepsilon_{\rm AA}/|\varepsilon_{\rm AB}| = -1$, and several values of pressure. 
Open circles determine a position of the critical points, broken lines display their pressure changes.  

\end{itemize}

\label{lastpage}

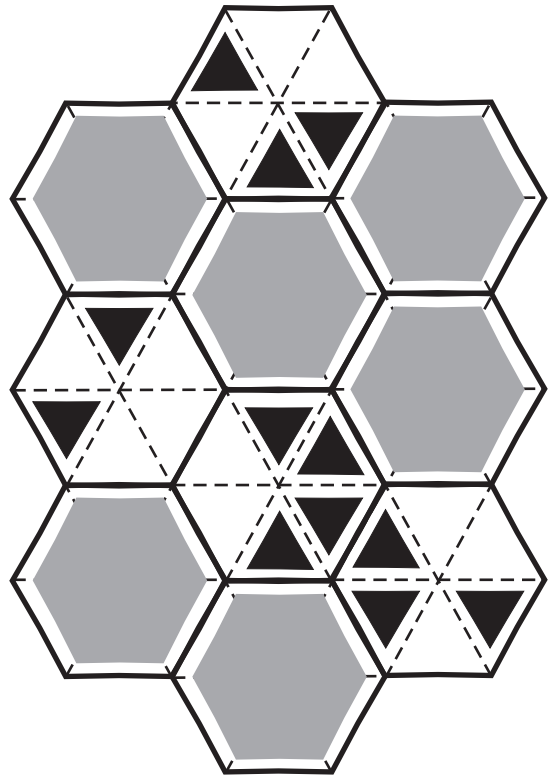
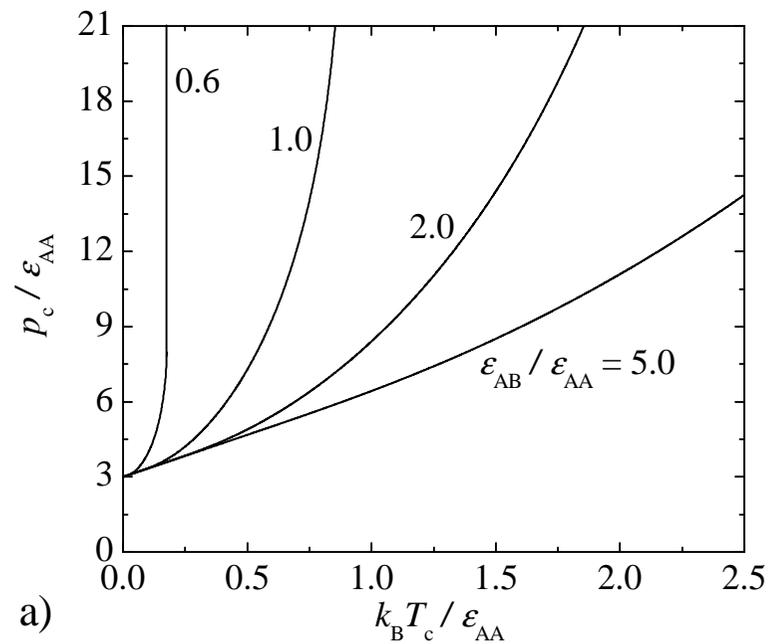
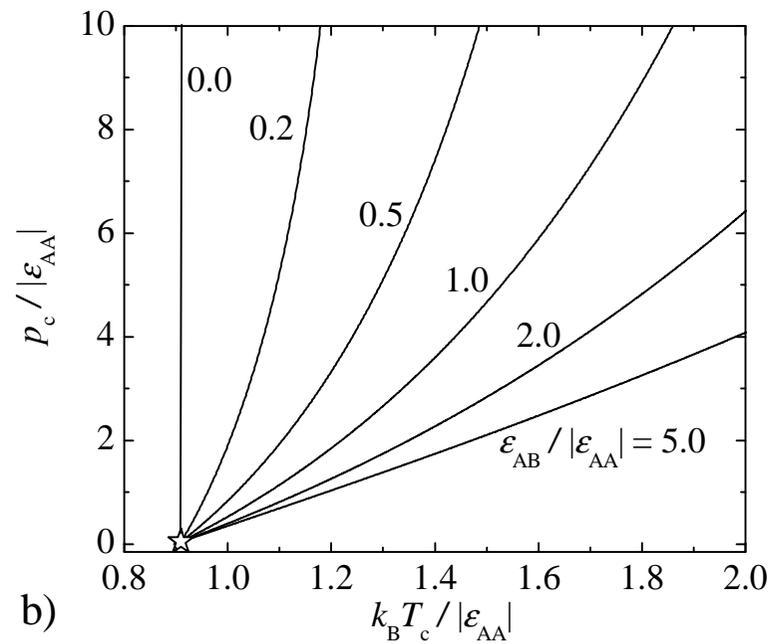
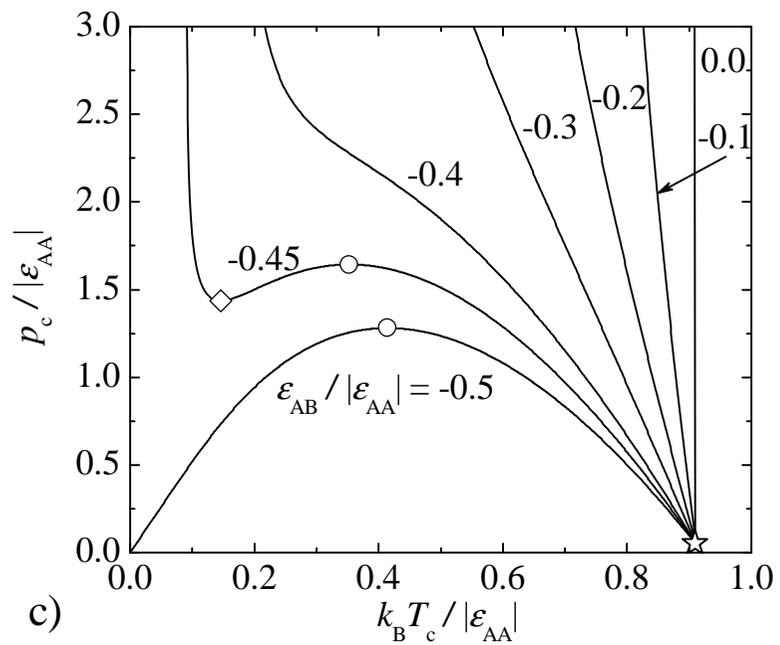
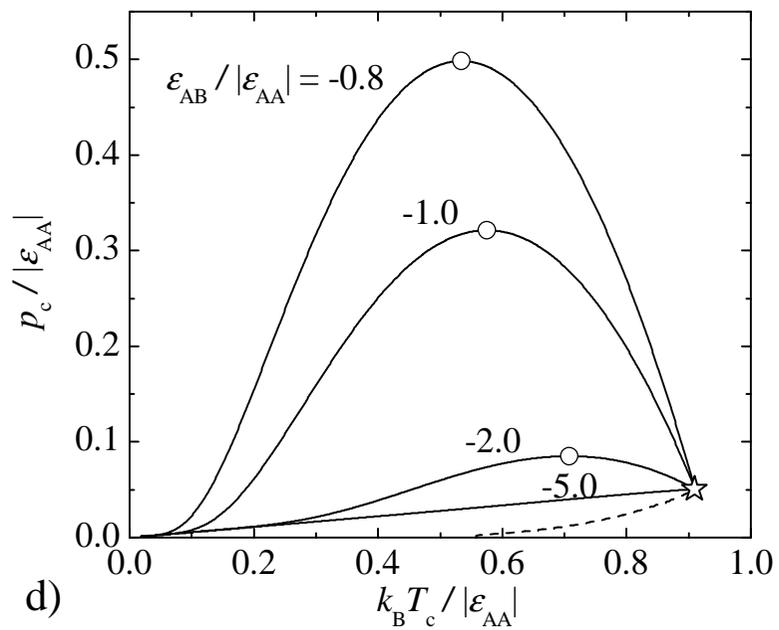
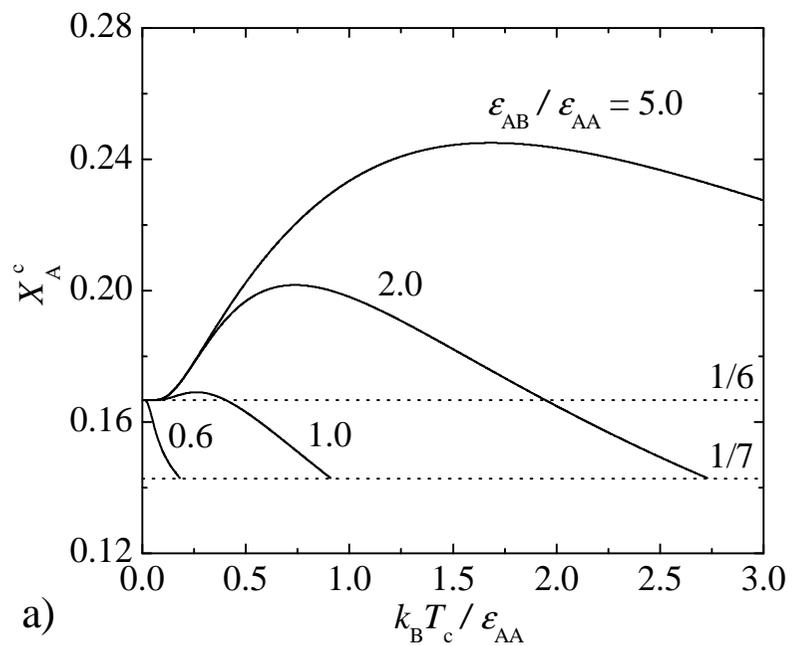
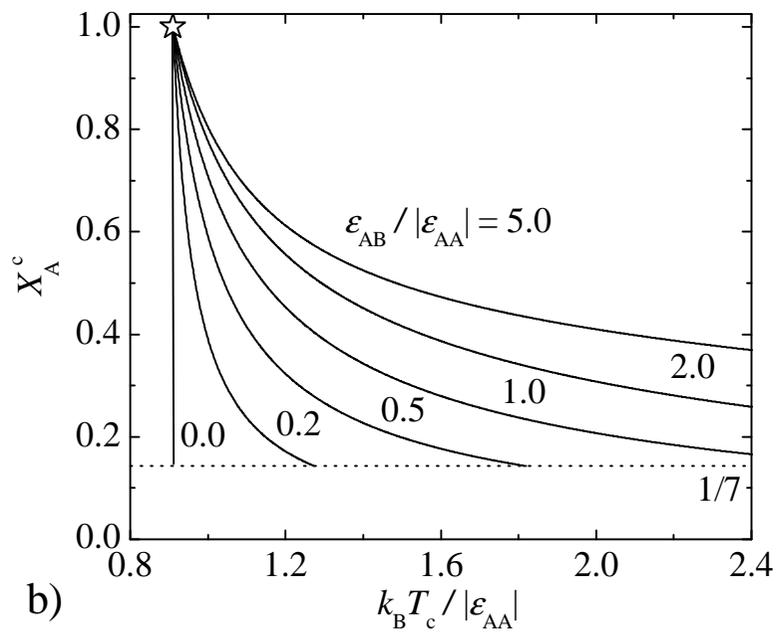
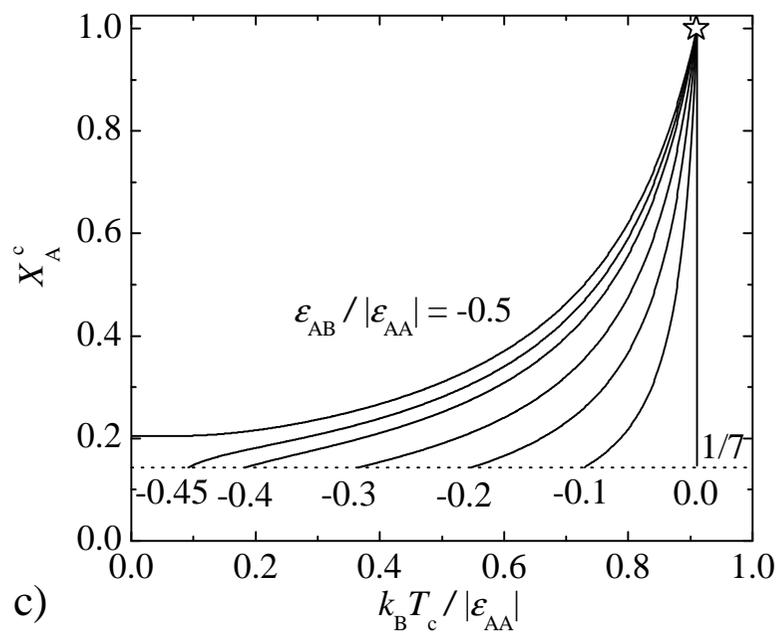
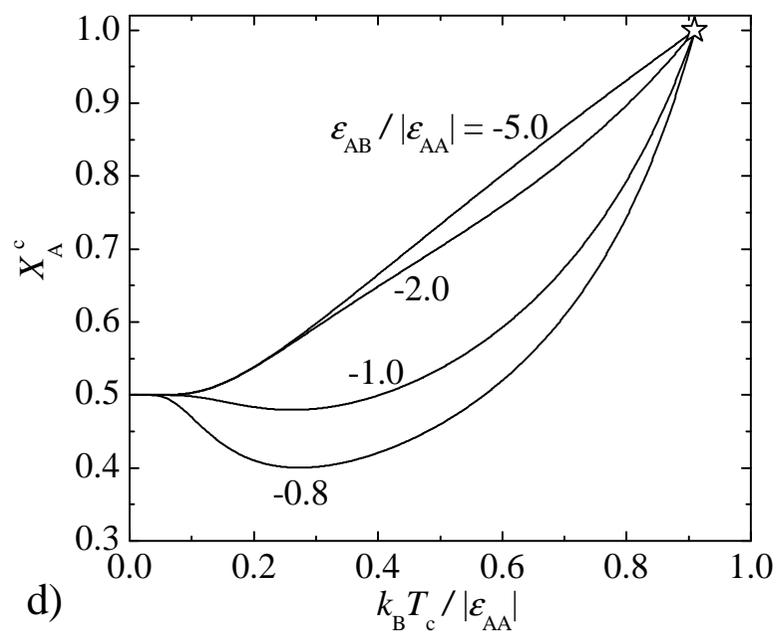
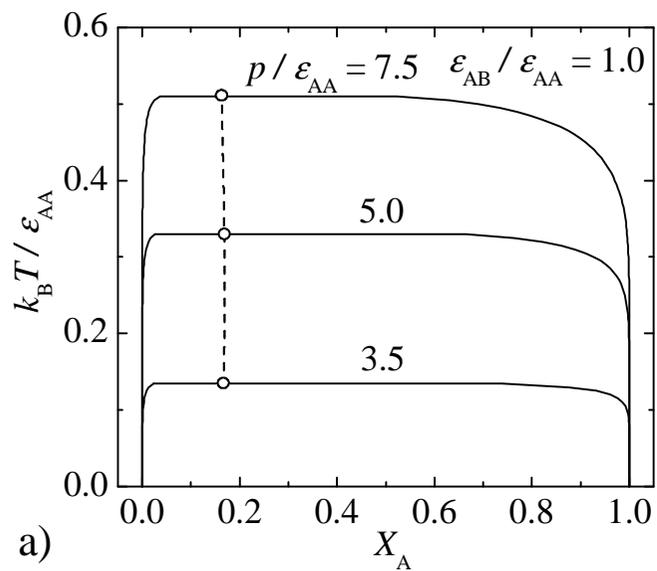
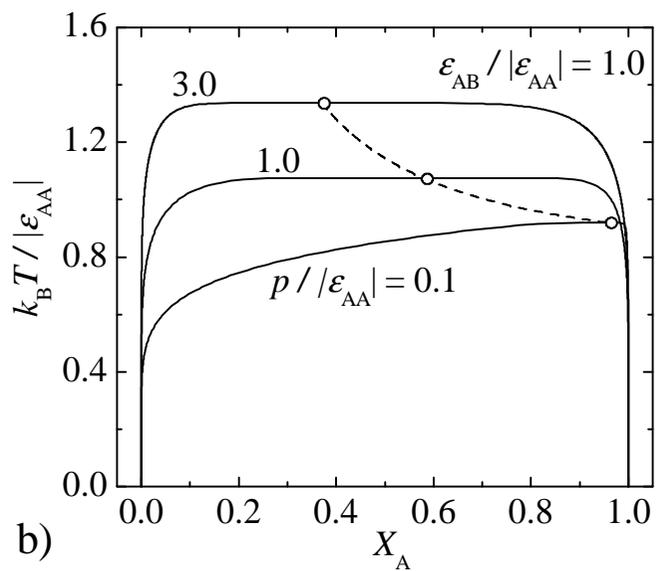
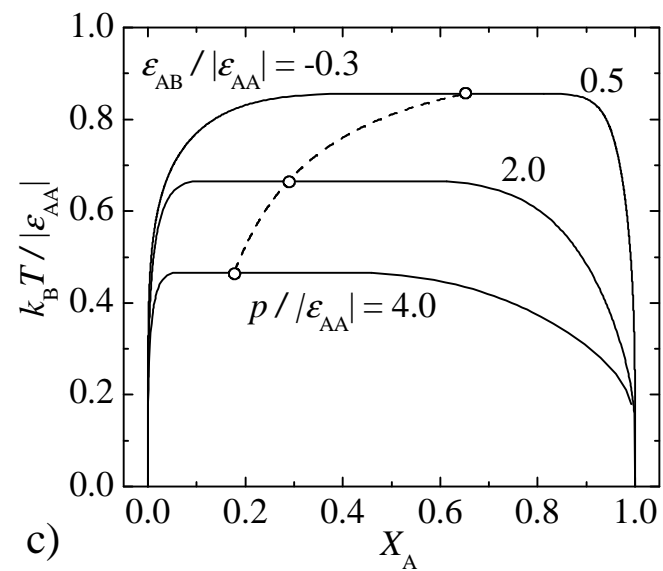
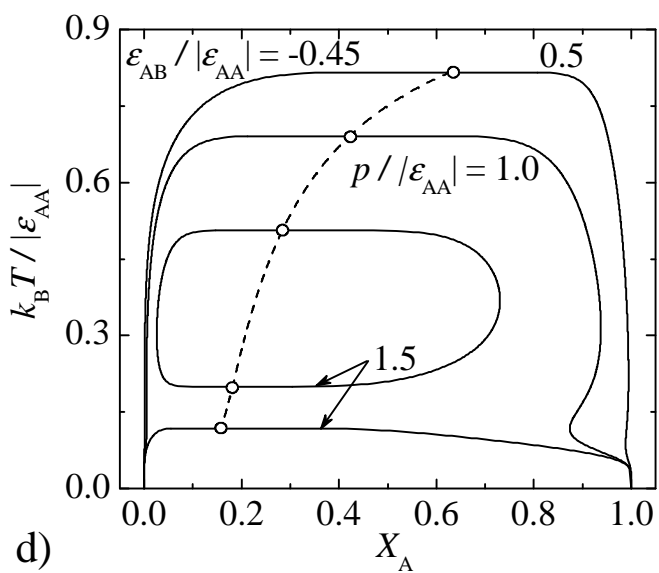
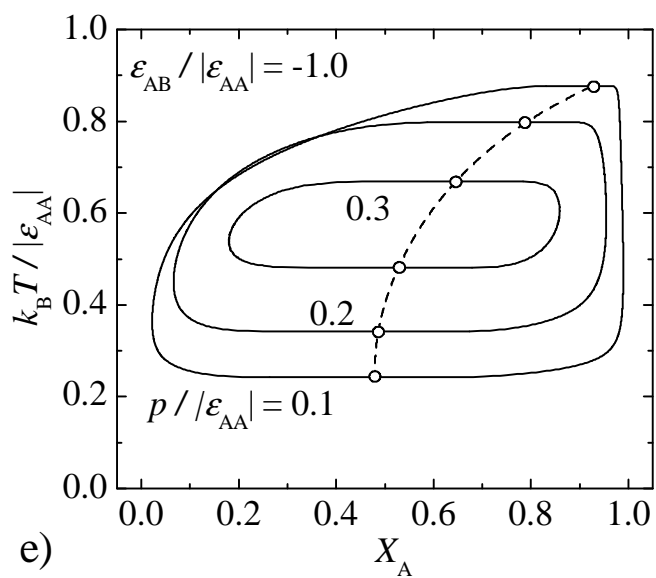
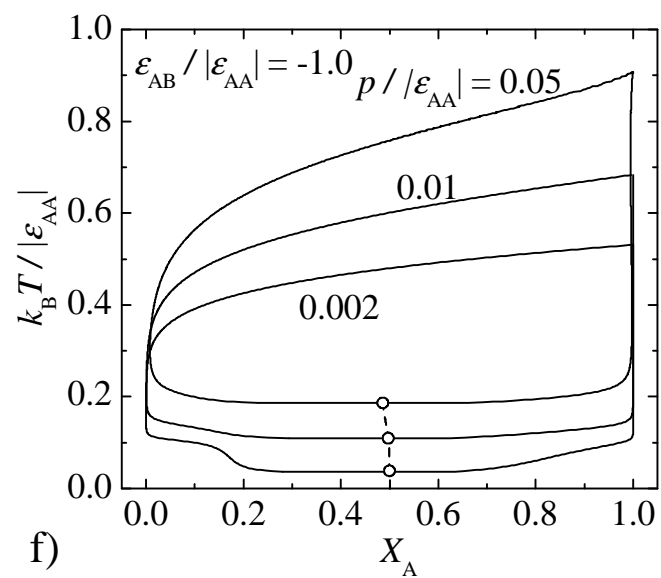
\end{document}